\documentclass[usenatbib]{mn2e}
\usepackage{amssymb}
\usepackage{color}
\usepackage{graphicx}

\newcommand{\be}{\begin{equation}}
\newcommand{\ee}{\end{equation}}
\newcommand{\ba}{\begin{eqnarray}}
\newcommand{\ea}{\end{eqnarray}}
\newcommand{\lp}{\left(}
\newcommand{\rp}{\right)}

\newcommand{\Pdot}{\dot{P}}
\newcommand{\Mdot}{\dot{M}}
\newcommand{\Omegas}{\Omega}
\newcommand{\OmegaK}{\Omega_{\mathrm{K}}}
\newcommand{\omegas}{\omega_{\mathrm{s}}}
\newcommand{\rmag}{r_{\mathrm{m}}}
\newcommand{\ralf}{r_{\mathrm{A}}}
\newcommand{\Nacc}{N_{\mathrm{acc}}}
\newcommand{\Nmag}{N_{\mathrm{m}}}
\newcommand{\Ntot}{N_{\mathrm{tot}}}
\newcommand{\Peq}{P_{\mathrm{eq}}}
\newcommand{\Bq}{B_{\mathrm{QED}}}
\newcommand{\mel}{m_{\mathrm{e}}}

\title[ESPs unite NS populations]{Equilibrium spin pulsars unite neutron star
 populations}
\author[Ho, Klus, Coe, \& Andersson]{Wynn C. G. Ho,$^{1,2}$\thanks{Email:
 wynnho@slac.stanford.edu}
H. Klus,$^{1,3}$ M. J. Coe$^{1,3}$ and Nils Andersson$^{1,2}$
\\
$^1$STAG Research Centre, University of Southampton, Southampton, SO17 1BJ \\
$^2$Mathematical Sciences, University of Southampton, Southampton, SO17 1BJ \\
$^3$Physics \& Astronomy, University of Southampton,
 Southampton, SO17 1BJ
}
\date{Accepted 2013 November 5. Received 2013 November 5;
 in original form 2013 September 19}

\begin{document}
\pagerange{\pageref{firstpage}--\pageref{lastpage}} \pubyear{2013}

\maketitle

\label{firstpage}

\begin{abstract}
Many pulsars are formed with a binary companion from which they can accrete
matter.
Torque exerted by accreting matter can cause the pulsar spin to increase
or decrease, and over long times, an equilibrium spin rate is achieved.
Application of accretion theory to these systems provides a probe of the
pulsar magnetic field.
We compare the large number of recent torque measurements of accreting
pulsars with a high-mass companion to the standard model for how accretion
affects the pulsar spin period.
We find that many long spin period ($P\gtrsim 100\mbox{ s}$) pulsars must
possess either extremely weak ($B<10^{10}\mbox{ G}$) or extremely strong
($B>10^{14}\mbox{ G}$) magnetic fields.
We argue that the strong-field solution is more compelling, in which case
these pulsars are near spin equilibrium.
Our results provide evidence for a fundamental link between pulsars with the
slowest spin periods and strong magnetic fields around high-mass companions
and pulsars with the fastest spin periods and weak fields around low-mass
companions.
The strong magnetic fields also connect our pulsars to magnetars and
strong-field isolated radio/X-ray pulsars.
The strong field and old age of our sources suggests their magnetic field
penetrates into the superconducting core of the neutron star.
\end{abstract}


\maketitle

\section{Introduction} \label{sec:intro}

The vast majority of the more than 2000 known neutron stars are observed in
the radio as rotation-powered pulsars, which lose rotational energy and
spin down due to their electromagnetic radiation.
Measurement of the rate of energy loss, or spin-down rate, allows one to
infer the pulsar magnetic field $B$ \citep{gunnostriker69,spitkovsky06}.
However the spin-down torque due to electromagnetic radiation
($\sim -10^{30}\mbox{ erg }B_{12}^2P^{-3}$, where $B_{12}=B/10^{12}\mbox{ G}$
and $P$ is the spin period)
is relatively weak compared to torque due to matter accretion by the neutron
star from a close binary companion.
In this latter case,
interactions between the pulsar magnetic field and matter from the companion
star produces a torque
($\sim\pm 10^{34}\mbox{ erg }\Mdot_{-9}^{6/7}B_{12}^{2/7}$, where
$\Mdot_{-9}=\Mdot/10^{-9}\mbox{ $M_\odot$ yr$^{-1}$}$ and $\Mdot$ is the
mass accretion rate) which can spin the pulsar up or down,
depending on whether matter is accreted or ejected.
By measuring the rate of change in spin period $\Pdot$
(which is related to torque $N$ by $N=-2\pi I\Pdot/P^2$, where $I$ is the
stellar moment of inertia),
one can obtain insights into the physics that governs the accretion process,
including important properties such as the role of the magnetic field.
However, observing the long-term $\Pdot$ in accretion-powered pulsars
is very difficult due to short-term fluctuations, with approximately twenty
sources detected (see \citealt{klusetal13a}, and references therein).

Our study is based on the work of \citet{klusetal13}, who report a
detailed analysis of Rossi X-ray Timing Explorer (RXTE) observations of
high-mass X-ray binaries (HMXBs) in the Small Magellanic Cloud,
specifically pulsar systems with an OBe main sequence companion;
these sources are designated Small Magellanic Cloud X-ray pulsars
(SXPs) \citep{coeetal05}.
The data set spans 13 years and contains 42
sources for which the spin period $P$ and count rate as a function of time
have been measured.  Spin period time derivatives $\Pdot$ are calculated
from the former, and X-ray luminosities $L$ are determined from the latter
and can be related to mass accretion rate by $\Mdot=LR/GM$, where $M$ and
$R$ are the neutron star mass and radius, respectively.
From their binary parameters, \citet{klusetal13} find that matter accretes
onto the neutron star in each SXP system via an accretion disk (c.f. wind).
Different methods are then used to determine the magnetic field of each pulsar.

Here we extend the work of \citet{klusetal13}.
In Section~\ref{sec:gl}, we examine torque models, especially that of the
standard disk accretion model of \citet{ghoshlamb79} and
\citet{kluzniakrappaport07}, and
demonstrate a simple method for matching the detailed result of
\citet{ghoshlamb79}.
In Section~\ref{sec:results}, we show that the standard model can explain
fast-spinning, weak magnetic field pulsars in low-mass X-ray binaries
(LMXBs) and slow-spinning, strong field pulsars in HMXBs, and
we consider the broader context of neutron star magnetic fields
in light of our findings.  We summarize in Section~\ref{sec:discuss}.

\vspace{-0.2cm}
\section{Standard Disk Accretion Model and Equilibrium Spin Pulsars}
\label{sec:gl}

\begin{figure}
\begin{center}
\includegraphics[scale=0.42]{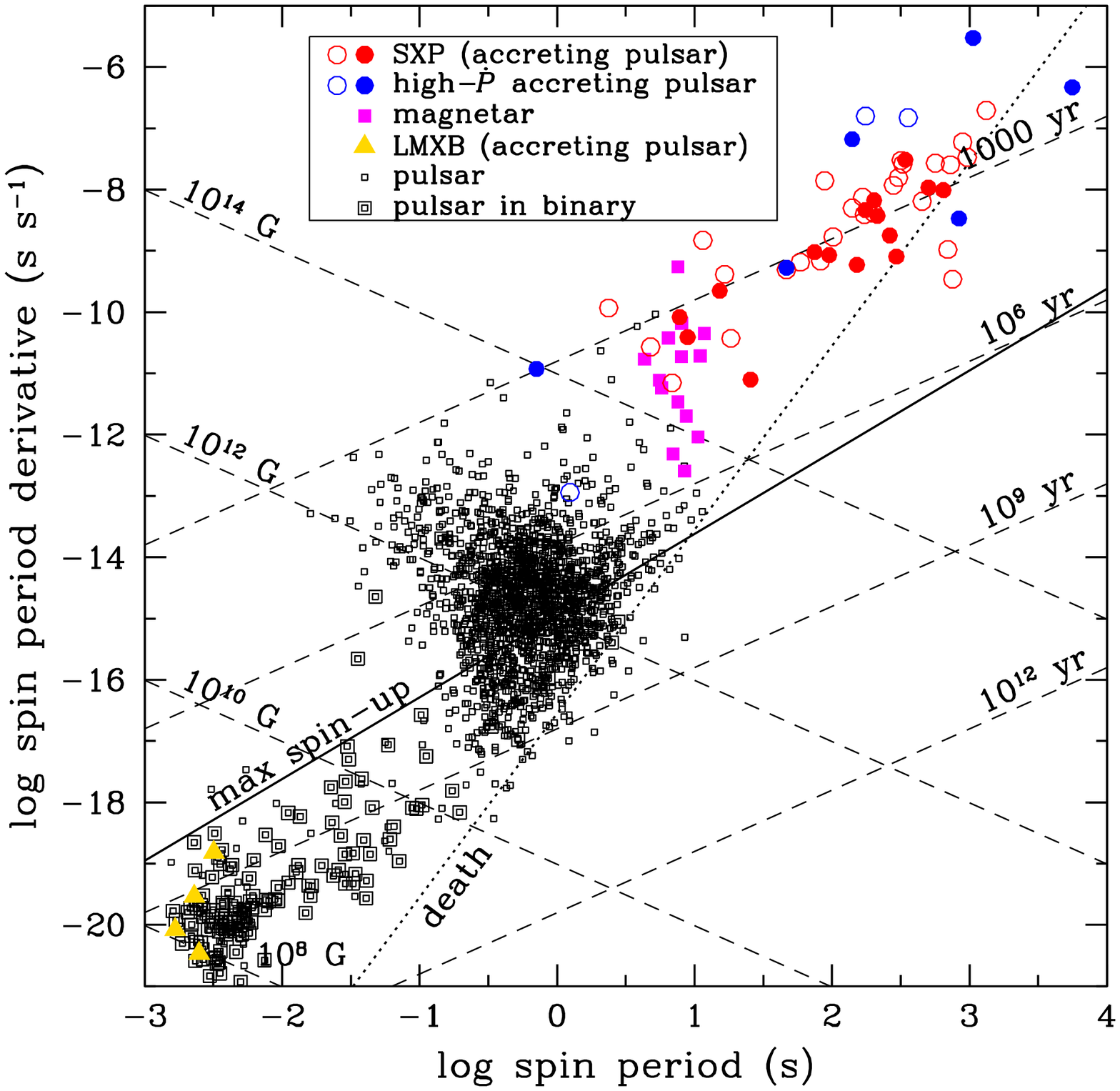}
\caption{
Pulsar spin period $P$ versus spin period time derivative $\Pdot$.
Open squares are pulsar values taken from the ATNF Pulsar Catalogue
\citep{manchesteretal05}, and solid squares denote magnetars.
Open and closed circles denote (accreting) sources that have $\Pdot<0$
and $\Pdot>0$, respectively (\citealt{klusetal13a}, and references therein;
\citealt{espositoetal13,klusetal13}).
Triangles denote (accreting) LMXBs
\citep{patruno10,haskellpatruno11,riggioetal11}.
The dashed lines indicate spin-down age ($=P/2\Pdot$) and inferred
magnetic field [$=3.2\times10^{19}\mbox{ G }(P\Pdot)^{1/2}$].
The dotted line indicates the (theoretically uncertain) death line
for pulsar radio emission;
note that the death line shown here is calculated using eq.~(\ref{eq:torqueem})
and therefore does not apply to accreting pulsars (c.f. Fig.~\ref{fig:pb}).
The solid line indicates the minimum $P$ and maximum $\Pdot$ that a pulsar
can possess as a result of matter accretion from a binary companion.
}
\label{fig:ppdot}
\end{center}
\vspace{-0.2cm}
\end{figure}

Figure~\ref{fig:ppdot} shows the spin period time derivative $\Pdot$ as a
function of spin period $P$ for known pulsars.
For isolated sources, i.e., those not in a binary system, pulsars with
the highest $\Pdot$ values are magnetars, i.e., neutron stars that
predominantly have magnetic fields $B\gtrsim 10^{14}\mbox{ G}$ and can
exhibit a variety of high-energy emission \citep{woodsthompson06,mereghetti08}.
The vast majority of sources are normal rotation-powered radio pulsars
whose spin-down rate (i.e., $\Pdot>0$) is measured very accurately from
coherent timing analysis (\citealt{lynegrahamsmith98}).
The pulsar magnetic field is then estimated by assuming that the
electromagnetic energy radiated produces a torque (after using the conversion:
torque~$=-2\pi I\Pdot/P^2$)
\be
\Pdot\approx 9.8\times 10^{-16}\mbox{ s s$^{-1}$ }R_6^{6}I_{45}^{-1}
 B_{12}^2P^{-1}, \label{eq:torqueem}
\ee
where $R_6=R/10\mbox{ km}$ and $I_{45}=I/10^{45}\mbox{ g cm$^2$}$.
While the exact nature of the mechanism that causes radio emission is not
known for certain, there is general agreement that there exists a
``death line'' below which observable emission
ceases \citep{rudermansutherland75,bhattacharyaetal92}.
An example death line is shown in Fig.~\ref{fig:ppdot}.

\begin{figure}
\begin{center}
\includegraphics[scale=0.42]{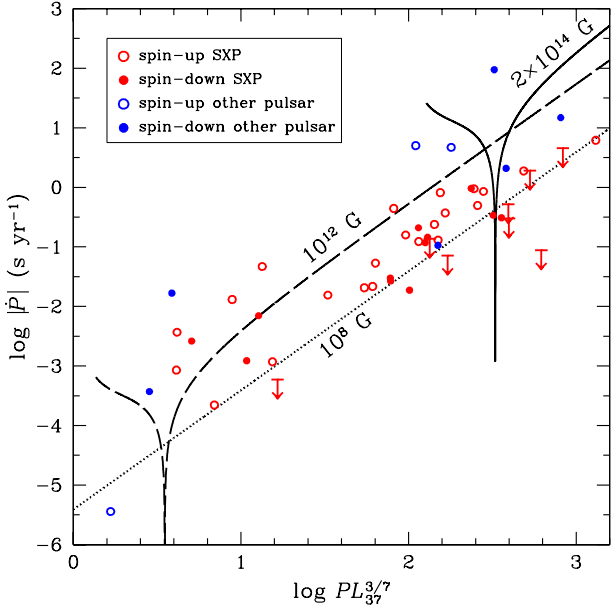}
\caption{
Rate of change of spin period $\Pdot$ versus $PL^{3/7}$.
Open and closed circles denote sources that have $\Pdot<0$ and $\Pdot>0$,
respectively (\citealt{klusetal13a}, and references therein;
\citealt{espositoetal13,klusetal13}).
The lines indicate the theoretical prediction from the standard disk accretion
model of \citet{ghoshlamb79} with different values of magnetic field
[see eq.~(\ref{eq:gl})].
}
\label{fig:gl}
\end{center}
\vspace{-0.2cm}
\end{figure}

In contrast to radio pulsars, the period derivative of accreting neutron
stars in an X-ray binary is determined by measuring and finding the
difference between the spin period at different epochs
(see, e.g., \citealt{townsendetal11}).
We see from Fig.~\ref{fig:ppdot} that $\Pdot$ for accreting pulsars
with $P\gtrsim 1\mbox{ s}$, such as the SXPs, is much larger than that
of most radio pulsars and that $\Pdot$ for SXPs is comparable to other
previously-known long spin period sources.
All these pulsars possess a binary companion (some companions are
low-mass stars and others are high-mass main sequence or supergiant stars)
from which the neutron stars are accreting.
Because accretion is thought to suppress radio emission
\citep{bhattacharyavandenheuvel91,archibaldetal09} and the torque from
accretion is much stronger than that of electromagnetic spin-down
(see Sec.~\ref{sec:intro}), the magnetic field of accreting pulsars
in a LMXB or HMXB cannot be estimated using eq.~(\ref{eq:torqueem}).
To determine their magnetic field, one can use the standard disk accretion
model of \citet{ghoshlamb79} (see also \citealt{ghoshlamb79a};
we also examine the model of \citealt{kluzniakrappaport07}, see below).
This model is based on detailed calculations of the interaction
between a rotating pulsar magnetosphere and an accretion disk surrounding
the pulsar.  The predicted torque yields
\ba
\Pdot &=& -4.3\times 10^{-5}\mbox{ s yr$^{-1}$ }
 M_{1.4}^{-3/7}R_6^{6/7}I_{45}^{-1} \nonumber\\
&& \times\, n(\omegas)\mu_{30}^{2/7}\lp PL_{37}^{3/7}\rp^2, \label{eq:gl}
\ea
where $M_{1.4}=M/1.4\,M_\odot$, $\mu_{30}=\mu/10^{30}\mbox{ G cm$^3$}$,
$\mu$ ($=BR^3$) is the magnetic moment of the neutron star,
and $L_{37}=L/10^{37}\mbox{ erg s$^{-1}$}$,
and is shown in Fig.~\ref{fig:gl} for three values of the magnetic field.
Hereafter we ignore mass and radius dependencies since they can only vary
by a factor of about two while the magnetic field can vary by several orders
of magnitude.
The dimensionless torque $n(\omegas)$ accounts for coupling between the
magnetic field and the disk plasma \citep{ghoshlamb79}
(see also \citealt{wang95}) and depends on the fastness parameter $\omegas$
[$\equiv\Omegas/\OmegaK(\rmag)$], which is given by
\be
\omegas = 3.3\,\xi^{3/2}M_{1.4}^{-2/7}R_6^{-3/7}\mu_{30}^{6/7}
 \lp PL_{37}^{3/7}\rp^{-1}. \label{eq:fastness}
\ee
$\Omega=2\pi/P$ is the pulsar spin frequency,
$\OmegaK(\rmag)$ is the Kepler orbital frequency at radius $\rmag$
[$=\xi\ralf=\xi(GM\mu^{4}/2R^2L^{2})^{1/7}$], where the energy density of
accreting matter transitions from being kinetically to magnetically
dominated, and $\xi\approx 1$ (see, e.g., \citealt{wang96}).
The sign of the dimensionless torque $n(\omegas)$ is determined by whether
the centrifugal force due to stellar rotation ejects matter and spins down
the star (``fast rotator'' regime with $\omegas>1$) or accretes matter and
spins up the star (``slow rotator'' regime with $\omegas<1$)
(see, e.g., \citealt{wang95}).
It is important to note that a long spin period ($P\gg 1\mbox{ s}$) pulsar
can still be classified as a fast rotator since the fastness parameter
$\omegas$ depends on the strength of the pulsar magnetic field.

\begin{figure}
\begin{center}
\includegraphics[scale=0.42]{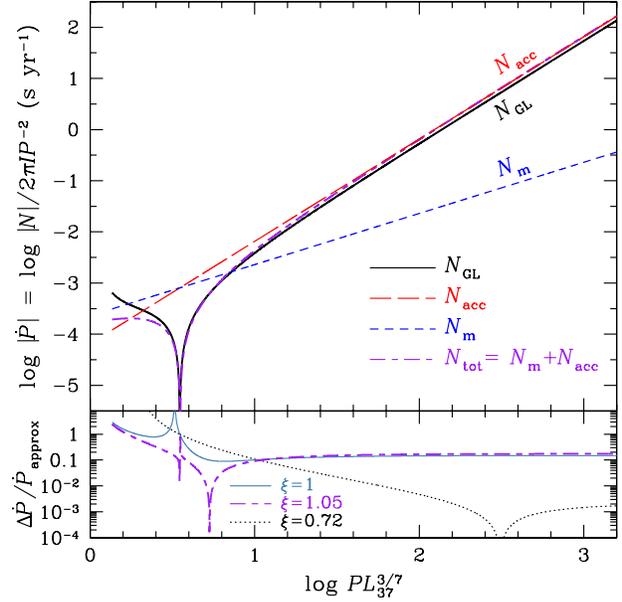}
\caption{
Top panel: Rate of change of spin period $\Pdot$ versus $PL^{3/7}$
for $B=10^{12}\mbox{ G}$.
The solid line labeled $N_{\mathrm{GL}}$ indicates the theoretical prediction
of \citet{ghoshlamb79} [see eq.~(\ref{eq:gl})].
The short-long-dashed line labeled $\Ntot$ indicates the analytic approximation
given by eq.~(\ref{eq:glapprox}).
Bottom panel: Relative difference between the exact and approximate values
of $\Pdot$ for different values of $\xi$ [see eq.~(\ref{eq:fastness})].
}
\label{fig:glapprox}
\end{center}
\vspace{-0.2cm}
\end{figure}

A good approximation to eq.~(\ref{eq:gl}) can be derived very simply.
Matter accreting at the magnetosphere transition $\rmag$ has specific
angular momentum
\be
l_{\mathrm{acc}}=\pm\rmag^2\OmegaK,
\ee
while matter rotating with the magnetosphere at $\rmag$ has specific
angular momentum
\be
l_{\mathrm{m}}=\rmag^2\Omegas.
\ee
Assuming prograde rotation between the accretion disk and neutron star
(i.e., $l_{\mathrm{acc}}>0$), the total torque on the neutron star is then
\be
\Ntot = \Nacc+\Nmag = \Mdot\Delta l = \Mdot\rmag^2\OmegaK\lp 1-\omegas\rp,
 \label{eq:torquetot}
\ee
which gives
\ba
\Pdot &=& -7.1\times 10^{-5}\mbox{ s yr$^{-1}$ }
 \xi^{1/2}M_{1.4}^{-3/7}R_6^{6/7}I_{45}^{-1} \nonumber\\
&& \times\, \lp 1-\omegas\rp \mu_{30}^{2/7}\lp PL_{37}^{3/7}\rp^2.
 \label{eq:glapprox}
\ea
Figure~\ref{fig:glapprox} shows that eq.~(\ref{eq:glapprox}) agrees with
eq.~(\ref{eq:gl}) to within $\approx 15\%$ over a large range in $PL^{3/7}$,
except when $\omegas$ approaches and exceeds unity.
Similar accuracy is obtained as the magnetic field is varied over many
orders of magnitude.

The linear dependence of $\log\Pdot$ at relatively high values of
$\log PL^{3/7}$
is simply the result of the standard {\it spin-up} torque
$\Mdot(GM\rmag)^{1/2}$ due to disk accretion
(for prograde rotation, or spin-down torque for retrograde rotation)
and is the first term $\Nacc$ in eq.~(\ref{eq:torquetot})
\citep{pringlerees72,rappaportjoss77}.
In the slow rotator regime [$\omegas\ll 1$; see eq.~(\ref{eq:fastness})],
the rate of change of spin period $\Pdot$ is simply given by $\Nacc$, and
the magnetic field is $B\propto \Pdot^{7/2}$.
The second term in eq.~(\ref{eq:torquetot}) is the {\it spin-down} torque
due to mass accretion onto a rotating object.
These two torques and their total (recall that $N=-2\pi I\Pdot/P^2$)
are shown in Fig.~\ref{fig:glapprox}.
We note that an equation similar to eq.~(\ref{eq:torquetot}) is
given in \citet{shakuraetal12} but with different coefficients for $\Nacc$
and $\Nmag$.

The sharp decrease and then increase in $\Pdot$ for $B=10^{12}\mbox{ G}$
(and $2\times 10^{14}\mbox{ G}$ in Fig.~\ref{fig:gl}) is due to the change of
sign from $\Pdot<0$ (spin-up) to $\Pdot>0$ (spin-down).
This marks the fast rotator regime near spin equilibrium ($\omegas\approx 1$)
\citep{davidsonostriker73,alparetal82,wang87,kluzniakrappaport07}:
After sustained epochs of spin-up and spin-down, a steady state will be
achieved where the net torque on the pulsar is negligible ($\Pdot\sim 0$).
The resulting equilibrium spin pulsar (ESP) spins near the period
\begin{equation}
\Peq = 23\mbox{ s }\xi^{3/2}M_{1.4}^{-2/7}R_6^{15/7}L_{37}^{-3/7}
\lp B/10^{13}\mbox{ G}\rp^{6/7}, \label{eq:spineq}
\end{equation}
which is obtained by setting $\omegas=1$ in eq.~(\ref{eq:fastness}).
Note that for spin equilibrium at $\omegas\sim 0.35$ \citep{ghoshlamb79},
the coefficient in eq.~(\ref{eq:spineq}) becomes 67~s, and the inferred
magnetic field is lower by a factor of $(0.35)^{7/6}\sim 0.3$.

\vspace{-0.2cm}
\section{Comparison to SXP observations} \label{sec:results}

Previously only $\sim 20$ X-ray pulsars have had large torques (i.e., $\Pdot$)
measured (see, e.g., \citealt{klusetal13a}).
Figure~\ref{fig:gl} shows these, as well as the 42 SXPs from \citet{klusetal13}.
Using the observed values of $P$, $\Pdot$, and $L$ for each SXP,
\citet{klusetal13} find that the standard disk accretion model prediction
for $\Pdot$ with magnetic fields in the range $B\approx 10^6-10^{15}\mbox{ G}$
can fit all SXPs
(the model of {Klu\'{z}niak} \& Rappaport yields magnetic fields
$\sim 10-20\%$ weaker).
In particular, fits to each SXP can yield two possible solutions, either a
weak field ($<10^{10}\mbox{ G}$) or strong field ($>10^{12}\mbox{ G}$).
The weak field solution arises from the match between the observed $\Pdot$
and the torque $\Nacc$ [c.f. $\Ntot$; see eq.~(\ref{eq:torquetot})] if the
pulsar is in the slow rotator regime (see also Fig.~\ref{fig:glapprox}).
On the other hand, if the SXP is in the fast rotator regime near spin
equilibrium, then the strong pulsar field is simply determined from
eq.~(\ref{eq:spineq}),
and the observed $\Pdot$ is due to small fluctuations around torque balance
($\Nacc\approx\Nmag$).
Both solutions are remarkable and unexpected, based on previous observations
of other types of pulsars (see Fig.~\ref{fig:pb}).
In the following, we argue for the strong magnetic field solution.
We note that, prior to the present work, only four pulsars have been found
to possibly be near spin equilibrium
\citep{patruno10,haskellpatruno11,riggioetal11},
although many pulsars in LMXBs are assumed to be in this regime
\citep{whitezhang97,patrunoetal12}.
The strong field solution we find (see also \citealt{klusetal13}) suggests
that a vast majority of the 42 SXPs are near spin equilibrium and thus are
ESPs.

\subsection{Magnetic Fields: Weak or Strong?} \label{sec:magb}

For many SXPs, we find that their observed $P$, $\Pdot$, and $L$ can be fit
with the standard accretion model [eq.~(\ref{eq:gl})] and one of two possible
magnetic fields: a relatively weak field ($<10^{10}\mbox{ G}$) or strong
field ($>10^{12}\mbox{ G}$).
While either solution is valid, the strong field is more likely from
primarily age/timescale arguments.
Most neutron stars are born with $B>10^{12}\mbox{ G}$,
which can be inferred from the lines of magnetic field and spin-down age
shown in Fig.~\ref{fig:ppdot}, as well as from population synthesis studies
\citep{fauchergiguerekaspi06,popovetal10}.
Therefore if SXPs have weak magnetic fields, their initial field either
decayed to the current level or has been buried below the surface.
The global magnetic field in neutron stars decays rather slowly and on the
timescale $\sim 10^6\mbox{ yr}$
\citep{haenseletal90,goldreichreisenegger92,glampedakisetal11},
while the maximum age of an OBe star is
$\sim 10^7\mbox{ yr}$.  Thus there is insufficient time
for the field to decay the many orders of magnitude required.
Burial of the field by accretion \citep{romani90} is unlikely in so many
SXPs because of the very short time over which a large amount of accretion
must take place \citep{chevalier89,geppertetal99}.
On the other hand, if SXPs have strong magnetic fields, then many are
currently near spin equilibrium,
which they can easily achieve given their short spin-evolution timescale
($P/|\Pdot|\sim 10^3\mbox{ yr}$) compared to their age.
Finally, spin equilibrium also allows spin-down ($\Pdot>0$) without
retrograde rotation.

\subsection{Connection to LMXBs and millisecond pulsars} \label{sec:lmxb}

The evolutionary scenario that leads to the formation of fast-spinning
($P\lesssim 10\mbox{ ms}$, hence millisecond) pulsars is briefly described
here (see, e.g., \citealt{alparetal82,bhattacharyavandenheuvel91,tauris12}).
A millisecond pulsar begins as a standard pulsar in a wide binary that
has spun down past the death line.
This pulsar can be brought back to radio activity by accreting
matter from its (low-mass) companion star, either when the orbit shrinks due
to gravitational radiation or the companion evolves off the main sequence.
Accretion of matter spins up the pulsar, thus recycling it back from
beyond death, although it is only seen as a millisecond radio pulsar after
accretion ceases.
However some pulsars are observed in the X-rays which are emitted during
the accretion process, and these are the LMXBs \citep{bildstenetal97}.
Thus LMXBs are predecessors of millisecond radio pulsars.
In this scenario, there is a lower limit to the spin period that
accretion can produce.
This minimum period is obtained by considering a pulsar that has reached
spin equilibrium through accretion at the maximum accretion rate or luminosity
$\sim 10^{38}\mbox{ erg s$^{-1}$}$,
above which radiation pressure prevents further accretion.
Equation~(\ref{eq:spineq}) then yields the magnetic field of the pulsar,
and these fields turn out to be low for LMXBs ($B\sim 10^8-10^9\mbox{ G}$).
After accretion ceases, the pulsar spins down by the standard electromagnetic
energy loss.
The resulting maximum spin-up line is shown in Fig.~\ref{fig:ppdot}.
We see that almost all observed pulsars which are likely to be recycled
are below this line and above the death line.

\begin{figure*}
\begin{center}
\includegraphics[scale=0.6]{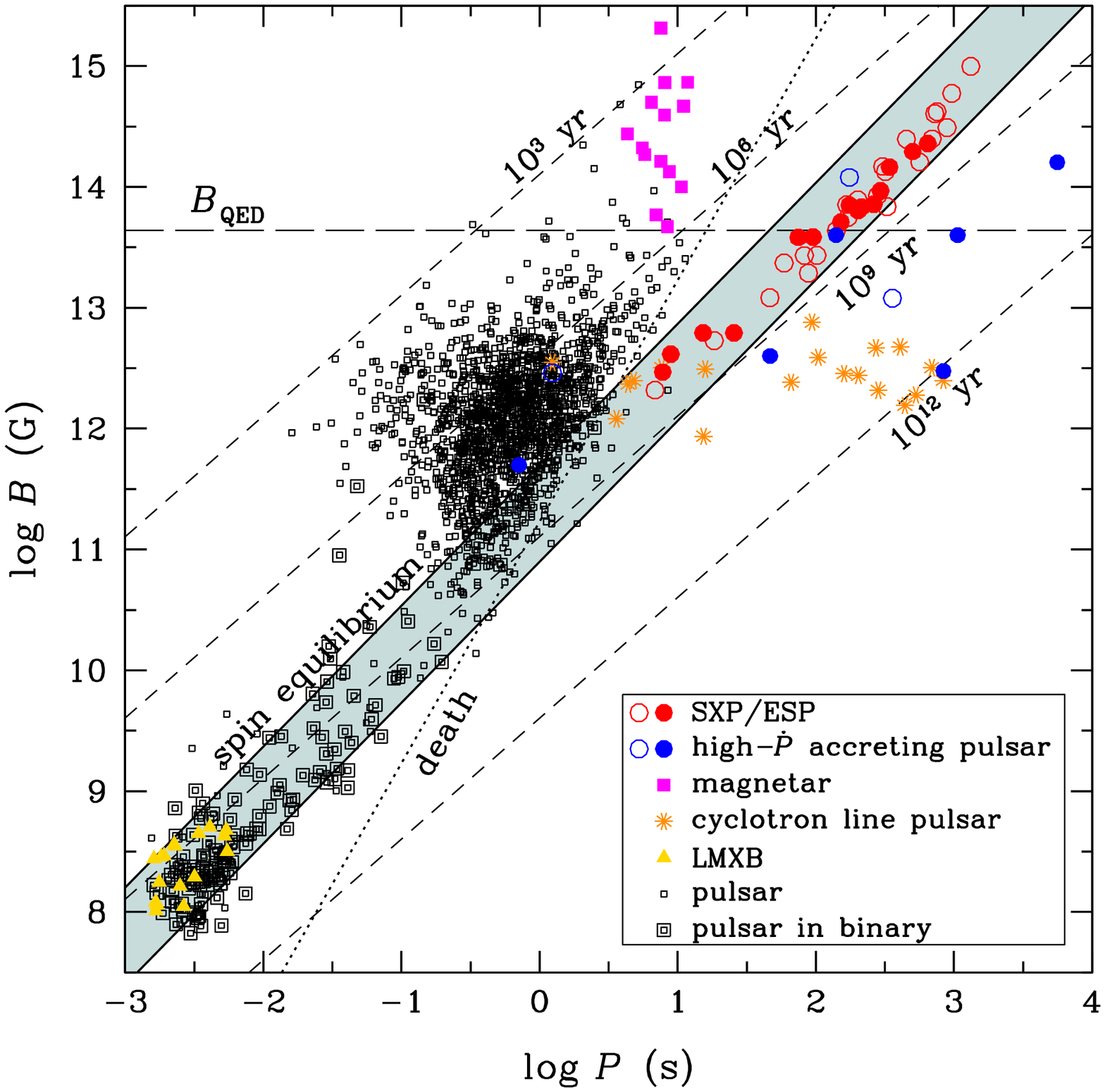}
\caption{
Pulsar magnetic field $B$ versus spin period $P$.
Open squares are pulsars from the ATNF Pulsar Catalogue
\citep{manchesteretal05}, and solid squares denote magnetars.
Open and closed circles denote sources that have $\Pdot<0$ and $\Pdot>0$,
respectively
\citep{ghoshlamb79,baykaletal02,cuismith04,ikhsanov12,reigetal12,klusetal13a,klusetal13}.
Triangles denote LMXBs \citep{patrunoetal12}.
Stars denote high mass X-ray binaries whose field is determined from an
electron cyclotron spectral line
(see http://www.sternwarte.uni-erlangen.de/wiki/doku.php?id=cyclo:start).
The short-dashed lines indicate spin-down age ($=P/2\Pdot$), the
long-dashed line indicates the QED field $\Bq=4.414\times 10^{13}\mbox{ G}$,
and the dotted line indicates the (theoretically uncertain) death line for
pulsar radio emission.
The solid lines bounding the shaded region indicate the $B$ and $P$
[see eq.~(\ref{eq:spineq})] that a pulsar can possess as a result of
accretion from a binary companion at the mass accretion rate observed
for SXPs/ESPs (red circles).
}
\label{fig:pb}
\end{center}
\vspace{-0.2cm}
\end{figure*}

Figure~\ref{fig:pb} shows the magnetic field and spin period of known pulsars.
The magnetic field of rotation-powered pulsars are obtained from their
$P$ and $\Pdot$ and using eq.~(\ref{eq:torqueem}).
Only four LMXBs have a measured $\Pdot$ or upper limit on $\Pdot$,
and spin equilibrium is assumed in order to calculate their magnetic field.
Many LMXBs have large distance errors which affects the determination of
their luminosity.
Thus their inferred magnetic fields are very uncertain, and this uncertainty
limits our ability to understand the physics of accretion using LMXBs.

The SXPs bridge this gap in knowledge.
The weak magnetic field solution to these pulsars would place them in the
extreme bottom right of Fig.~\ref{fig:pb}, in a region far away from all
previously known sources, and thus is not shown.
Instead we show the strong magnetic field solution, which is determined
from their measured spin period and X-ray luminosity and
eq.~(\ref{eq:spineq}) since SXPs are near spin equilibrium for this solution.
The shaded region in Fig.~\ref{fig:pb} is the spin equilibrium region for the
range of luminosities [$L=(0.2-4)\times 10^{37}\mbox{ erg s$^{-1}$}$]
spanned by our sources.
Evidently SXPs (as ESPs) are governed by the same accretion physics as that
used to explain the recycling of LMXBs into millisecond pulsars even though
the physics regime spans many orders of magnitude in magnetic field.
While LMXBs, with their low magnetic fields ($B\sim 10^8-10^9\mbox{ G}$),
produce pulsars with millisecond spin periods,
SXPs, with their high magnetic fields ($B\sim 10^{12}-10^{15}\mbox{ G}$),
produce pulsars with long spin periods ($P\gtrsim 10\mbox{ s}$).
An important difference between the LMXBs and SXPs is that the latter
are now known to be close to spin equilibrium, as well as having a more
accurate distance, and thus their magnetic fields are much better determined.

\subsection{Connection to magnetars and high-$B$ pulsars} \label{sec:magnetars}

The strong magnetic field of SXPs spans a wide range
($B\approx 10^{12}-10^{15}\mbox{ G}$),
similar to the range for rotation-powered pulsars.
There are also $<20$ Galactic HMXBs for which electron cyclotron spectral
lines have been detected; however there is uncertainty in where these lines
are generated and likely selection effects (see \citealt{klusetal13}, for
more details).
Of particular note are those 24 of 42 SXPs with
$B>\Bq=\mel^2c^3/e\hbar=4.414\times 10^{13}\mbox{ G}$,
where $\Bq$ is the critical quantum electrodynamics (QED) magnetic field
(and 13 SXPs have $B>10^{14}\mbox{ G}$).
There are about two dozen previously known neutron stars
($\sim 1\%$ of the total population) with these fields. They are composed
of magnetars, which are strong X-ray/gamma-ray sources and can undergo
transient outbursts both of which are powered by their strong magnetic
fields \citep{woodsthompson06,mereghetti08},
and high-$B$ pulsars, which behave similar to the bulk of the radio
pulsars \citep{ngkaspi11}.
The difference between these two groups could be due to the strength of
internal toroidal fields \citep{ponsperna11}.
Magnetar-like behavior has not been seen in SXPs, which indicates SXPs
with $B>\Bq$ are in the latter group and possess weak toroidal fields.
This may suggest that formation of high-$B$ pulsars is more efficient than
that of magnetars and there exists many more of the former than latter.

The relative number of SXPs with superstrong fields ($B>\Bq$) is much higher
than in the normal pulsar population.
On the one hand, this significantly higher fraction of highly-magnetized
neutron stars could be due to inherently different source populations,
e.g., neutron star formation as a result of an electron capture supernova
\citep{nomoto84,podsiadlowskietal04} or accretion induced collapse
\citep{nomoto84,taamvandenheuvel86,nomotokondo91}.
On the other hand, selection effects could be hindering detection of
isolated highly-magnetized neutron stars.
Effects include X-ray absorption, radio dispersion, and/or quenching of
radio emission beyond the death line,
and some of these effects are taken into account in population synthesis
models (see, e.g., \citealt{fauchergiguerekaspi06,popovetal10}).
Note that for SXPs, their rotational energy loss rate $\dot{E}$
[$=4\pi^2I\Pdot/P^3$ and using eq.~(\ref{eq:torqueem})] is
$\sim 4\times 10^{25}\mbox{ erg s$^{-1}$}
(B/10^{13}\mbox{ G})^2(P/100\mbox{ s})^{-4}$.
In addition, magnetars and high-$B$ pulsars have a short spin-down
timescale ($\approx 10^3-10^5\mbox{ yr}$; see Figs.~\ref{fig:ppdot} and
\ref{fig:pb}) and are
young (age on the same order as their spin-down timescale).
Meanwhile, the high-mass companion of SXPs has a much longer lifetime
($\sim\mbox{a few}\times 10^6\mbox{ yr}$),
which allows us to study the pulsars at later stages in their life.
In other words, it is only because of accretion that we are able to detect
these highly-magnetized neutron stars when they are older than when they are
young and isolated (see also \citealt{chashkinapopov12}).
This may suggest that magnetic field decay in neutron stars occurs slower
than previously thought \citep{ponsetal09}
and could be the result of the magnetic field extending into the
superconducting neutron star core, where the field decay timescale is
much longer \citep{glampedakisetal11}.
Importantly, this scenario where all pulsars are drawn from the same
source population would resolve the neutron star birthrate
problem \citep{keanekramer08,wattersromani11} and
support a unified picture of neutron stars
\citep{keanekramer08,kaspi10,popovetal10}.

\vspace{-0.2cm}
\section{Summary} \label{sec:discuss}

Using recent measurements of a large number of spin period time derivatives
$\Pdot$ for pulsars in high-mass binaries in the Small Magellanic Cloud
\citep{klusetal13}, we examined the torque implications for the standard disk
accretion model \citep{ghoshlamb79} and model of \citet{kluzniakrappaport07}
and for neutron star magnetic fields.
We find several interesting results.
Many of the 42 SXPs are near spin equilibrium, where torques that increase
and decrease the spin period are balanced, and thus SXPs are ESPs;
previously only four pulsars have been found to possibly be near spin
equilibrium.
The standard disk accretion model links SXPs (which have high mass companion
stars) with pulsars at very short spin periods (which have low mass
companions); this is demonstrated by the shaded region in Fig.~\ref{fig:pb}.
The strong magnetic field of many SXPs is above the QED-field and links
them with magnetars and strong-$B$ radio/X-ray pulsars.
The decay of magnetic field in SXPs occurs after $10^6$~yr, suggesting
magnetic field penetration into the superconducting pulsar core.
It is possible there are many more superstrong magnetic field pulsars in
the Galaxy that remain as yet undetected.

\vspace{-0.2cm}
\section*{acknowledgments}
We thank the anonymous referee for helpful comments.
WCGH and NA acknowledge support from the Science and Technology
Facilities Council (STFC) in the United Kingdom.
HK acknowledges a STFC studentship.

\bibliographystyle{mnras}

\label{lastpage}

\end{document}